\documentclass{emulateapj}
\usepackage{textcomp}
\usepackage{mathtext}
\usepackage{amssymb}

\newcommand{\fermi}{\textit{Fermi}}
\newcommand{\gr}{$\gamma$-ray}

\begin{document}

\title{Supernova Remnant Kesteven 27: Interaction with A Neighbor HI 
Cloud Viewed by \textit{Fermi}}

\author{Yi Xing\altaffilmark{1}, Zhongxiang Wang\altaffilmark{1},
Xiao Zhang\altaffilmark{2}, \& Yang Chen\altaffilmark{2,3}}

\altaffiltext{1}{\footnotesize 
Key Laboratory for Research in Galaxies and Cosmology,
Shanghai Astronomical Observatory, Chinese Academy of Sciences,
80 Nandan Road, Shanghai 200030, China}

\altaffiltext{2}{\footnotesize Department Astronomy, Nanjing University,
22 Hankou Road, Nanjing 210093, China}

\altaffiltext{3}{\footnotesize 
Key Laboratory of Modern Astronomy and Astrophysics,
Nanjing University, Ministry of Education, Nanjing 210093, China}

\begin{abstract}
We report on the likely detection of $\gamma$-ray emission from
the supernova remnant (SNR) Kesteven 27 (Kes~27). We 
analyze 5.7 yr \fermi\ 
Large Area Telescope data of the SNR region and find
an unresolved source at a position consistent with the radio 
brightness peak and the X-ray knot of Kes 27, which is located in 
the eastern region of 
the SNR and caused by the interaction with a nearby HI cloud.
The source's emission is best fit with a power-law spectrum with a photon index 
of 2.5$\pm$0.1 and a $>$0.2 GeV luminosity of 
5.8$\times 10^{34}$ erg s$^{-1}$ 
assuming a distance of 4.3 kpc, as derived from radio observations of 
the nearby HI cloud. 
Comparing the properties of the source with that 
of other SNRs that are known to be interacting with nearby
high-density clouds, 
we discuss the origin of the source's emission.
The spectral energy distribution of the source can be described by a hadronic
model that considers the interaction of energetic protons, escaping from 
the shock front of Kes~27, with a high-density cloud.
\end{abstract}


\section{Introduction}

The high sensitivity and fine spatial resolution 
of \textit{the Fermi Gamma-ray Space telescope},
along with that of the ground-based very high energy (VHE) \gr\ telescopes, have
allowed us to conduct unprecedentedly detailed study of supernova 
remnants (SNRs) at high-energy GeV and TeV energies. 
From \fermi\ observations, we now know that due to their interaction 
with nearby molecular clouds (e.g., \citealt{fs2012}),
middle-aged SNRs, such as W51C \citep{abdo+2009}, W44 \citep{abdo+1-2010}, IC 443 \citep{abdo+2-2010}, and W28 \citep{abdo+3-2010}, 
are among the brightest MeV to GeV \gr\ sources, having $\gamma$-ray 
luminosities of $\sim$10$^{36}$ erg~s$^{-1}$. As a comparison,
young SNRs with ages no larger than a few thousands of years, 
such as RX J1713.7$-$3946 \citep{eps+2010,aaa+2011-1}, 
Cas A \citep{abdo+CasA2010}, and Tycho \citep{aaa+2011}, 
have $\gamma$-ray luminosities two orders of magnitude lower.
$\gamma$-ray emission from dynamically evolved SNRs 
interacting with molecular clouds
is believed to be dominated by pion decay emission,
resulting from collision of relativistic protons with ambient material 
(e.g., \citealt{abdo+2009,abdo+1-2010,abdo+2-2010,abdo+3-2010}).
Their high luminosities rule out the alternative 
leptonic origin, since the required total electron energy would 
be $>$10$^{51}$~erg, larger than the typical kinetic energy 
released by a supernova explosion (e.g., \citealt{abdo+3-2010}).
In the leptonic scenario, the high energy emission is thought to 
be due to either inverse Compton up-scattering of ambient low-energy 
photons by relativistic electrons or Bremsstrahlung radiation 
from high-energy electrons.
Even among young ($\lesssim$2000 yrs) SNRs, sometimes there is also 
evidence suggesting a hadronic 
origin to the observed $\gamma$-ray emission. The position of \gr\ emission 
may coincide with the region in an 
SNR that is known to be interacting with a nearby molecular 
cloud (e.g., \citealt{xw2014}). Here in this paper, we report 
\fermi\ detection of another such case, the SNR Kesteven 27 (Kes 27).

As a thermal composite SNR, Kes 27 (also known as G327.4$+$00.4) was 
found to be confined in an HI shell and interacting with a nearby HI cloud
on the southeast \citep{mgd+2001}, 
leading to a radio brightness peak at the southeast edge of 
Kes 27 \citep{mck+1989,mgd+2001}. 
The extent of the SNR's radio emission is approximately 
$\sim$21\arcmin\, as measured with the Molonglo Observatory Synthesis 
Telescope (MOST) at 843 MHz \citep{wg96}.
A kinematic distance of approximately 4.3$\pm$0.5 kpc was estimated 
for the HI cloud and thus Kes 27, derived from the velocity measurements from
the HI absorption spectra toward them \citep{mgd+2001}.
Kes 27 has also been observed at X-ray energies with different X-ray telescopes
including \textit{Einstein} \citep{lm1981,s1990}, \textit{ROSAT} \citep{skr1996}, \textit{ASCA} \citep{etm+2002,kon+2005}, and \textit{Chandra} \citep{css+2008}. 
Slightly different from other thermal composite SNRs whose X-ray emission is 
centrally peaked, high-spatial \textit{Chandra} imaging has revealed that 
this SNR has bright emission in the region east of its 
center and the emission near the eastern shell roughly coincides with 
the radio morphology \citep{css+2008}. The X-ray imaging thus also indicates 
the enhanced emission due to the interaction with the HI cloud. 
From X-ray observations of the diffuse emission from Kes 27, 
a dynamical age of $\sim$8000 yr was derived for the SNR, adopting a shock 
velocity of 580 km s$^{-1}$ and a diameter of 20\arcmin\ \citep{css+2008}.

In this paper we report our analysis of the \fermi\ Large Area Telescope
(LAT) data
of the Kes 27 region and the likely detection of \gr\ emission from
the interaction region of the SNR.
We describe the \fermi\ observation data we used in Section~2,
and present the data analyses and results in Section 3.
The results are discussed in Section 4.

\section{Observation}
\label{sec:obs}

As the main instrument onboard the \textit{Fermi Gamma-ray Space 
Telescope}, LAT is a $\gamma$-ray imaging 
instrument that scans the whole sky every three hours and can conduct long-term \gr\ observations of sources in the energy range from 20 MeV to 300 GeV \citep{aaa+2009}.
In our analysis 
we selected LAT events from the \textit{Fermi} Pass 7 Reprocessed (P7REP) 
database inside a $\mathrm{20^{o}\times20^{o}}$ region 
centered at the position of Kes 27. 
The SIMBAD position of Kes 27 is R.A.$=237\fdg1583$, Decl.$=-53\fdg7867$ 
(equinox J2000.0), which was adopted as the central position of Kes 27 after 
comparing with the radio \citep{mck+1989} and X-ray 
maps \citep{skr1996,etm+2002,css+2008} of this source.
We kept events during the time period from 2008-08-04 15:43:36 (UTC) to 
2014-04-13 22:13:17 UTC, and rejected events below 200 MeV 
because of the relative large uncertainties of the instrument response 
function of the LAT in the low energy range. 
In addition, following the recommendations of the LAT team\footnote{\footnotesize http://fermi.gsfc.nasa.gov/ssc/data/analysis/scitools}, we included those
events with zenith angles less than 100 degrees, which 
prevents the Earth's limb contamination, and 
during good time intervals
when the quality of the data was not affected by the spacecraft events. 

\section{Analysis and Results}

\subsection{Source Detection}
\label{subsec:si}

We included all sources within 16 degrees centered at the position of 
Kes 27 in the \textit{Fermi} 2-year catalog \citep{naa+2012}
to make the source model. The spectral function forms of these 
sources are provided in the catalog. 
The spectral normalizations of the sources 
within 8 degrees from Kes 27 were set as free parameters, and 
the other parameters were fixed at their catalog values. 
In addition, the $\gamma$-ray pulsar PSR J1543$-$5149 \citep{nbb+2014} was
included in the source model, which was not listed in the catalog. 
We modeled the pulsar's emission with a power law with an exponential cut-off, which is the characteristic spectrum for pulsar emission, and set 
the spectral normalization, spectral index, and cutoff energy as free
parameters.  The Galactic and extragalactic diffuse 
emission was also added in the source model with the spectral model 
gll\_iem\_v05.fits and the file iso\_source\_v05.txt, respectively, used.
The normalizations of the diffuse components were free parameters.

\begin{figure}
\epsscale{1.0}
\plotone{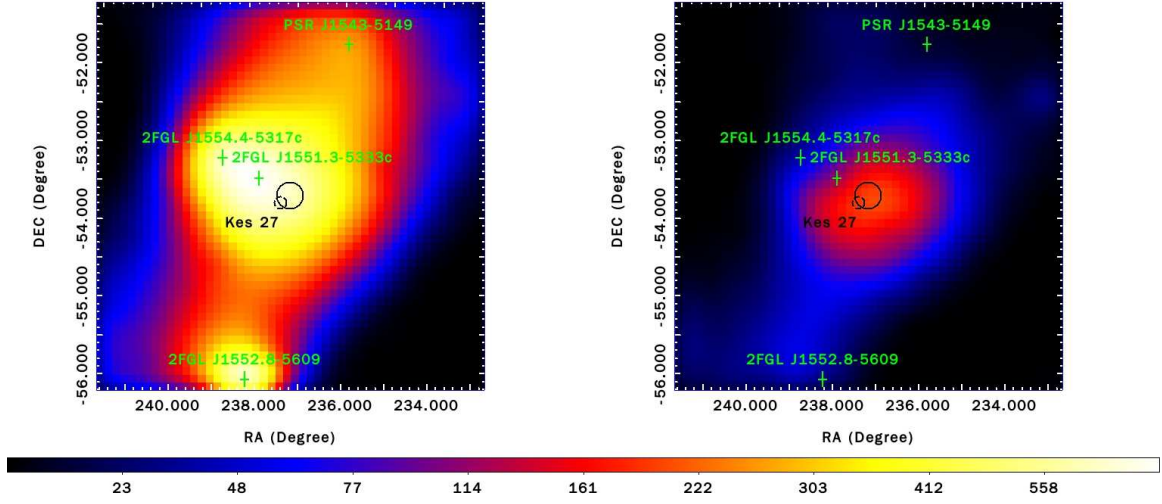}
\caption{TS maps (0.2--300 GeV) of the $\mathrm{5^{o}\times5^{o}}$ region 
centered at R.A. $=$ 237\fdg1583, Decl. $=$ $-$53\fdg7867 
(equinox J2000.0). The image scale of the maps is 0\fdg1 pixel$^{-1}$. 
The sources in the source model outside of the regions were considered 
and removed in the {\it left} panel, and all sources in the source model 
were considered and removed in the {\it right} panel. 
The green crosses mark the sources in the source model, the dark solid 
circle indicates the radio counterpart of Kes 27 \citep{g2009}, and 
the dark dashed circle marks the 2$\sigma$ error circle of the best-fit 
position of the Kes 27 $\gamma$-ray emission.}
\label{fig:lmap}
\end{figure}

We performed standard binned likelihood analysis to the LAT data 
in the $>$0.2 GeV range using the LAT science tools software 
package {\tt v9r23p5}, and extracted the Test Statistic (TS) 
map of a $5\arcdeg\times 5\arcdeg$ region centered 
at the position of Kes 27.  A TS map that included sources 
in the source 
model outside of the region was made, which is shown in the left 
panel of Figure~\ref{fig:lmap}.  
The TS map indicates that Kes 27 is located in a complex region with 
two nearby catalog sources, 2FGL J1554.4$-$5317c and 2FGL J1551.3$-$5333c. 
Then removing all the sources in the source model in 
this region, a residual map was made and is shown in the right 
panel of Figure~\ref{fig:lmap}.  Excess $\gamma$-ray emission 
appears near the center with TS$\simeq$217, indicating $\sim$15$\sigma$ 
detection significance. 
For these analyses, we also tested to included sources within 
20 degrees centered at Kes 27 and free the spectral indices of the sources 
within 5 degrees from Kes 27, but the results of source positions and spectra
did not have significantly differences (consistent within uncertainties) and
at the Kes 27 region, TS$\simeq 200$, only slightly lower than that
from fixing the spectral indices.

While \citet{lande12} analyzed the two nearby sources 
2FGL J1554.4$-$5317c and 2FGL J1551.3$-$5333c, and determined that they
did not have extended emission, we also investigated whether or not 
the excess emission could be due to any confusion because of 
the proximity to 2FGL J1551.3$-$5333c. Different TS maps in the energy
ranges of $>$1\,GeV, $>$2\,GeV, and $>$3\,GeV were made, as the point spread functions (PSFs) of
LAT are significantly reduced at the high energies\footnote{\footnotesize http://www.slac.stanford.edu/exp/glast/groups/canda/lat\_Performance.htm}. 
For making the TS maps, all the sources except 2FGL J1551.3$-$5333c
in the source model were removed. 
We found that the excess emission is clearly resolved from 
2FGL J1551.3$-$5333c. A TS map of the source region in 3--300\,GeV is 
displayed in the left panel of Figure~\ref{fig:hmap},  
showing that they are separate.
Actually at the high energy range, the catalog position 
of 2FGL J1551.3$-$5333c, 
having TS$\simeq$34, is not at the TS peak ($\simeq$48). There may be
another source east to 2FGL J1551.3$-$5333c. 
We tested to determine the position for this
source in 3--300\,GeV and obtained
R.A.=238\fdg22, Decl.=$-$53\fdg61, (equinox J2000.0), with a large
uncertainty of 0\fdg26 (1$\sigma$). The position is $\simeq$0\fdg2 away 
from 2FGL J1551.3$-$5333c, and therefore they are consistent within 
the uncertainty. No conclusion could be made for them
based on the current data. In any case, the excess emission at the
Kes 27 region was always present in these analyses.


Examining the excess emission at the high energy range,
we found that it actually consists of two individual sources.
The $1\arcdeg\times 1\arcdeg$ residual TS map of 
the region centered at Kes 27 (right panel of Figure~\ref{fig:hmap}),
which was made from using the $>$3\,GeV data,
shows the details. 
Two possible sources are resolved, with the southeast and
northwest ones marked as $A$ and $B$, respectively.
The TS values for them are $\simeq$22 and $\simeq$14, respectively. 
While source $B$ dominates the emission 
in the $>$0.2 GeV energy range (see \S~\ref{subsec:sa}), 
source $A$ is more significantly detected 
in the $>$3 GeV energy range. 
We overlaid the radio intensity contours for Kes 27, detected by 
MOST at 843 MHz \citep{wg96}, on the map, and found that
the radio brightness peak of Kes 27 is
located close to the southeast source.
By running \textit{gtfindsrc} in the LAT 
software package, we determined its position, which is 
R.A.=237\fdg37, Decl.=$-$53\fdg88, (equinox J2000.0), with 1$\sigma$ nominal 
uncertainty of 0\fdg04.
This position is consistent with that of the radio brightness peak 
of Kes 27 (approximately R.A.=237\fdg33, Decl.=$-$53\fdg82)
within the 2$\sigma$ error circle. The positional coincidence suggests
that source $A$ is likely associated with Kes 27. 
Below we considered this source as the \gr\ counterpart to Kes 27
(see also the Discussion section).

For source $B$, we also obtained its best-fit position in the $>$3 GeV 
energy range, and the position is R.A.=236\fdg86, Decl.=$-$53\fdg67, 
with 1$\sigma$ nominal uncertainty of 0\fdg08. 
We tested to consider and remove this source from the $>$3 GeV TS map, 
and the excess $\gamma$-ray emission (i.e., source $A$) at 
the southeast region of Kes 27 was still present with TS$\simeq$20. 

We thus added these two sources at their best-fit positions to the source 
model and performed binned likelihood analysis in the $>$0.2 GeV range. 
The spectra of them were modeled with a power law. 
We found that sources $A$ and $B$ have
photon indices of $\Gamma=2.5\pm$0.1 (with a TS value of $\sim$76)
and $\Gamma=2.7\pm$0.1  (with a TS value of $\sim$80), respectively.

Source B could be emission from Kes 27 too as it positionally
coincides with the northwest part of Kes 27 (Figure~\ref{fig:hmap}). 
However, given the radio
and X-ray morphology of the SNR, no notable features were seen
at the source's position, and it would be hard to explain why the SNR is
bright at this source's region but not in the whole region 
if the SNR has significant \gr\ emission.
In the \textit{Chandra} observation reported by \citet{css+2008},
an X-ray point source, CXOU J154816.7$-$534125, was detected at 
the position of R.A.=237\fdg0700, Decl.=$-$53\fdg6904 ($\sim$0.5\arcsec\ 
uncertainty). This source (marked by a green cross in the right panel
of Figure~\ref{fig:hmap}) is 0\fdg12 away from the best-fit
position of source $B$ but within the 2$\sigma$ error circle.
It was thought to be a background candidate
AGN. As AGNs generally have a power-law \gr\ spectrum, with photon index
up to 3.0 in the \fermi\ \gr\ energy range \citep{1stagn}, the spectrum of 
source $B$ also suggests the possible association between them
(for source $B$'s spectrum, see \S~\ref{subsec:sa}).

\subsection{Spatial Distribution Analysis}
\label{subsec:sda}

We analyzed the spatial distribution of the 
Kes 27 counterpart to determine whether it is point-like or 
extended. 
We used both a point source and uniform disk models with power-law spectra
at the best-fit position to analyze the emission in the 3--300 GeV range. 
Source $B$ was included in the source model. The searched radius range
for the uniform disks was 0\fdg1--0\fdg5 with a step of 0\fdg1. 
Additionally in the analysis, 
only front converting events for the instrument response function 
P7REP\_SOURCE\_V15::FRONT were included, which allows to reduce 
the PSF of the LAT point sources to $<$0\fdg3 (68\% containment) 
in $>$3 GeV range. 
For the point source, we set the spectral normalizations of the 
sources within 8 degrees from Kes 27 as free parameters, 
and fixed all the other parameters in the source model at 
the \fermi\ 2-year catalog values. The spectral indices of Kes 27 
and source $B$, 
and the spectral index and cutoff energy of PSR J1543$-$5149 were fixed at 
the values obtained from likelihood analysis in the $>$0.2~GeV energy range.  
For the disk models, we fixed all spectral parameters of the sources 
in the source model at the values obtained above, but set the spectral 
normalization parameters of the disk models free.
No significant extended emission was detected; TS$_{ext}$ values,
calculated from ${\rm TS}_{disk}-{\rm TS}_{point}$, were smaller than 0.

\subsection{Spectral Analysis}
\label{subsec:sa}

Considering the Kes 27 counterpart and source $B$ as 
point sources at their best-fit positions,
their $\gamma$-ray spectra were extracted by
performing maximum likelihood analysis to the LAT data 
in 10 evenly divided energy bands in logarithm from 0.1--300 GeV.
By assuming a power law for emission in each energy band,
the obtained fluxes are less model dependent, providing a good
description for the \gr\ emission of a source.
The source model included all sources in the \fermi\ 2-year 
catalog and the pulsar J1543$-$5149.
The spectral normalizations of the 
sources within 8 degrees from Kes 27 were set as free parameters, 
while all the other parameters of sources were fixed at the values 
we obtained above in Section~\ref{subsec:si}. 
We kept only spectral flux points with TS greater than 4 (corresponding to 
the detection significance of 2$\sigma$), and derived 95\% flux upper limits 
in the other energy bands. 
The obtained spectra of the counterpart to Kes 27 and source $B$ 
are shown in Figure~\ref{fig:spectra}, and the flux and uncertainty values
are given in Table~\ref{tab:tab1}.
From the flux measurements, it can be noted 
that source $B$ was more significantly detected at energies of $\lesssim$1~GeV.

In addition to the statistical uncertainties obtained above, we
note that the LAT effective area introduces approximately 5\%--10\% systematic uncertainties 
to the energy fluxes \citep{naa+2012,aaa+13}.
There are also systematic uncertainties due to the Galactic diffuse
emission model, which can be estimated from
repeating the likelihood analysis in each energy band, with the 
normalization of the diffuse component artificially fixed to the $\pm$6\% deviation
from the best-fit value (see e.g., \citealt{abdo+2009,abdo+3-2010,abdo+w49b2010}).
The uncertainties estimated in this way are provided in Table~\ref{tab:tab1}. 
They are the dominant ones in
the systematic uncertainties \citep{abdo+3-2010,aaa+13}.

\section{Discussion}
\label{sec:dis}

\subsection{Source Identification}

Having analyzed 5.7 year \fermi/LAT data of the Kes 27 region,
we found a $\gamma$-ray source at the position consistent with that of 
the radio brightness peak of the SNR. The $\gamma$-ray source has power-law 
emission with photon index of 2.5, and the total 0.2--300 GeV flux is
approximately 2.6$\times 10^{-11}$ erg~s$^{-1}$\,cm$^{-2}$. 
We performed spatial distribution analysis in the $>$3 GeV energy range, 
but no significant extended emission was detected for the source. 
The $>$0.2 GeV detection significance of the source for
a point-source profile at the best-fit position is $\simeq$8$\sigma$.

The positional coincidence strongly suggests the association of the
\gr\ source with Kes~27, the emission of which is
enhanced and thus detectable due to 
the SNR's interaction with a nearby dense cloud.
The \fermi\ detected SNRs that are interacting with dense clouds 
can appear to have prominent flux peaks around $\sim 1$\,GeV, which
is likely to be explained by the hadronic scenario (e.g., \citealt{lc2012} 
and references therein): a cloud with high mass density acts as a large
target for relativistic protons to interact with and decay into neutral 
pions and subsequently $\gamma$-rays.
For example, the SNRs W44 \citep{giu+2011}, 
HB 21 \citep{pht+2013}, and IC 443 \citep{aaa+2013} were observed to 
have such features in their spectra. 
Some of the interacting SNRs may not appear to have
the prominent features, with the spectra described by models such as a flat 
power law ($\Gamma\sim 2$; e.g., G296.5$+$10.0, \citealt{a2013}; W41, 
\citealt{csc+2013}; RCW 103, \citealt{xw2014}), a relative soft power law 
($\Gamma\geq 2.4$; e.g., MSH 17$-$39, \citealt{csc+2013}), or a curved power 
law (e.g., W51C, \citealt{abdo+2009}; Kes 79, \citealt{asc2014}). 
However, the spectral data points of these SNRs at energies of several hundreds of MeV were 
detected with low TS values and thus large uncertainties. When 
the spectral energy distributions (SEDs) of them were constructed 
combining results from observations at TeV energies, the SEDs were found 
to also peak around $\sim$1 GeV. 

The spectrum of the source at the southeast of Kes 27
is very similar in this respect
by having only significant emission above 1~GeV. Although 
this source is located in a complex region,
having two nearby sources (2FGL J1554.4$-$5317c and 2FGL J1551.3$-$5333c)
and an additional $\gamma$-ray source possibly associated with a background 
candidate AGN, the analysis of $>$3~GeV data clearly separates it
from the nearby sources. Further considering distance $d=4.3$\,kpc to 
the source, 
its luminosity is approximately 5.8$\times 10^{34} d^{2}_{4.3}$\,erg\,s$^{-1}$,
which is in the luminosity range of the currently detected SNRs.
The $\gamma$-ray luminosities are $\sim$10$^{33}$--10$^{34}$ erg s$^{-1}$ 
and $\sim$10$^{36}$ erg s$^{-1}$ for young and middle-aged SNRs, respectively,
although we note that several young SNRs with relatively hard $\gamma$-ray 
spectra 
probably have leptonic dominated emission (e.g., \citealt{aaa+2011-1,tan+2011}).

We searched in the SIMBAD Astronomical Database within 
the 2$\sigma$ error circle of the best-fit position of
the $\gamma$-ray source, and only a few normal stars are 
known in the region. 
The position, spectrum, and flux (or luminosity if we assume 4.3 kpc
source distance) of the source all support
its association with Kes 27. We thus conclude that we 
have likely found the \gr\ emission from the strongest interaction
region of Kes 27. 


\subsection{Origin of the $\gamma$-ray Emission}

\textit{Chandra} X-ray imaging of the field resolved the SNR into 
different substructures, one of which is a
dense knot in eastern side region E1 that
corresponds to the radio brightness peak. 
Emission from these substructures results from thermal plasma 
radiation, which does not provide any direct information about non-thermal 
particles at the shock (see \citealt{css+2008} for the details).
Radio flux measurements of
Kes 27 can be described by a power law with a spectral index of $-0.6$ 
\citep{ccc1975,mck+1989}, but are for the whole SNR.
We therefore only considered the \gr\ emission in the following modelling.

We first explored the possibility that the $\gamma$-rays originate 
from direct collision of the accelerated protons with the surrounding dense 
gas (the HI shell and, specifically, the southeastern cloud) without 
diffusive process. The detailed algorithm 
given in \citet{kel+06} was adopted here to calculate the 
$\pi^{0}$-decay $\gamma$-rays. We assumed that the energy distribution 
of accelerated protons have power law form with index $\alpha_{p}$, 
and that the average gas density of the surrounding target gas ($n_t$) 
is of order 1--10\,cm$^{-3}$. To explain the observed GeV fluxes, 
the energy converted into relativistic protons is required to be 
$W_p (> 1 {\rm GeV}) \sim 6\times10^{50} (n_t/1\, {\rm cm}^{-3}) ^{-1}$ erg. 
This corresponds to a fraction of the explosion energy converted into 
protons' energy  $\eta = 60\%(n_t/1\, {\rm cm}^{-3}) ^{-1} E_{51}^{-1}$ 
(where $E_{51}$ denotes the supernova explosion energy in units of 
$10^{51}$~erg). 
This is plausible in view of the presence of the surrounding dense HI gas 
and especially, the dense clump on the southeastern edge 
\citep{mgd+2001,css+2008}.  If the hot gas density 
(as high as $\ga 2$\,cm$^{-3}$) obtained from the X-ray emission along 
the eastern boundary (regions E1 and E2) can be a reference value for 
the target gas density, the conversion fraction $\eta$ would be in 
a moderate range within $30\%$.  
In this scenario, the proton index $\alpha_{p}$ is the same as the
$\gamma$-ray photon index, $2.5 \pm 0.1$, as obtained from the spectral fit. 
This value seems slightly high, as compared to 2.1--2.4 derived from 
the observed slope of the detected cosmic ray spectrum at Earth \citep{gab13}, 
and would be difficult to be theoretically explained for the shock accelerated 
protons.

\begin{figure}
\epsscale{0.55}
\plotone{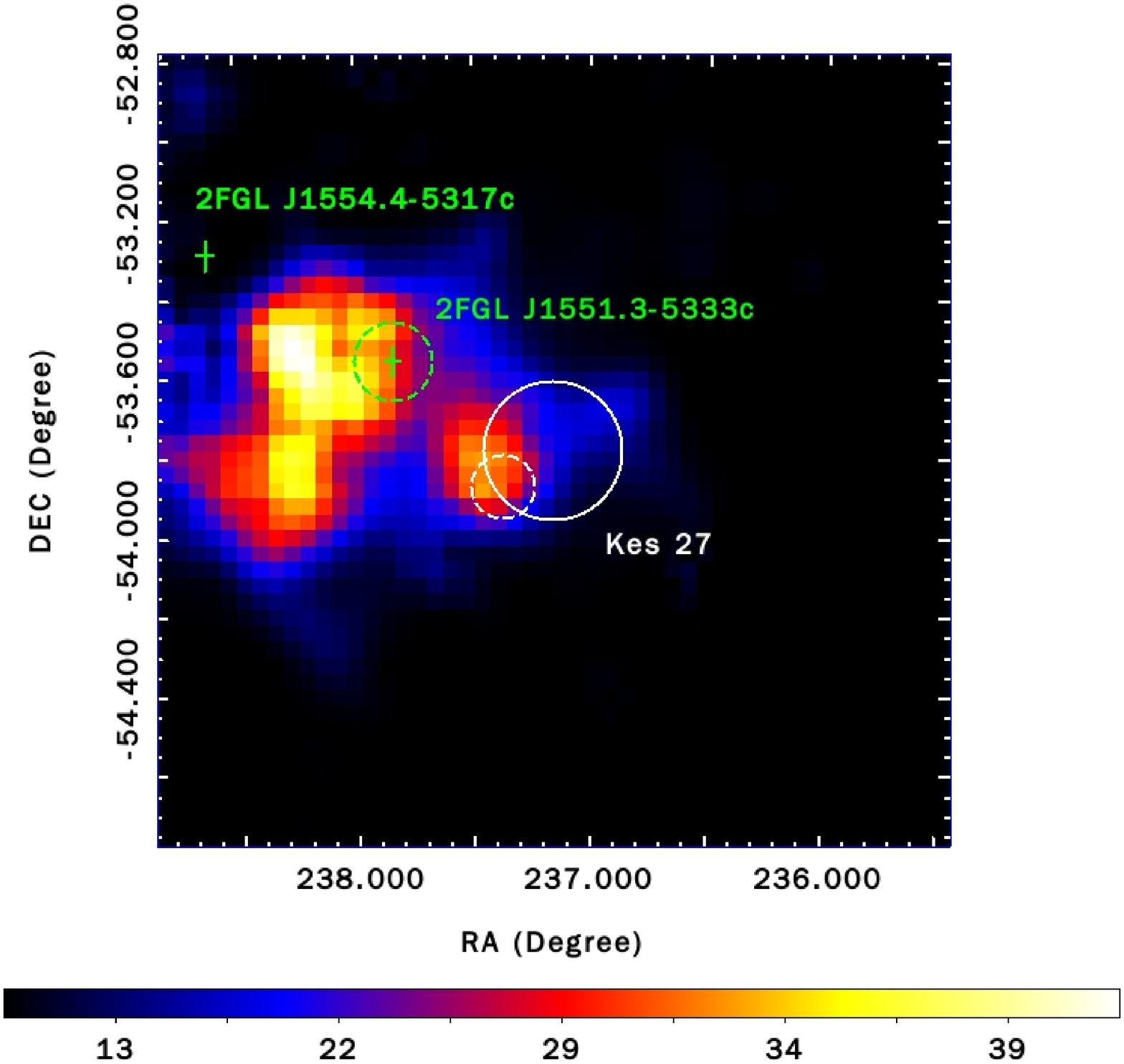}
\plotone{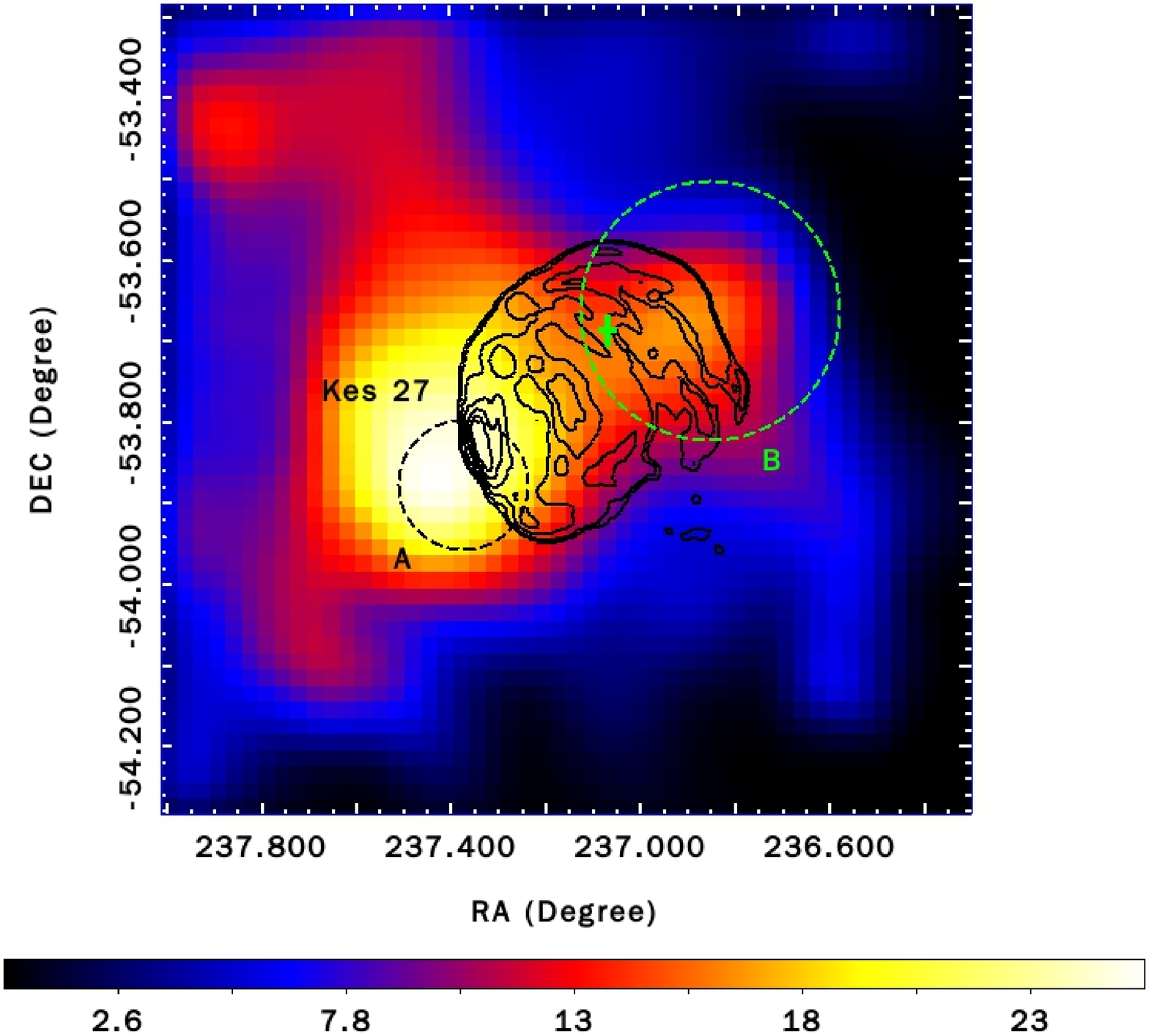}
\caption{{\it Left} panel: TS map of the $\mathrm{2^{o}\times2^{o}}$ region
centered at R.A. $=$ 237\fdg1583, Decl. $=$ $-$53\fdg7867 (equinox J2000.0) in the 3--300 GeV range. The image scale of the map 
is 0\fdg04 pixel$^{-1}$. All sources except 2FGL J1551.3$-$5333c were 
considered and removed. The symbols are the same as those 
in Figure~\ref{fig:lmap}, while the 2$\sigma$ error circle for the
nearby catalog source 2FGL~J1551.3$-$5333c is marked by a green dashed circle. 
{\it Right} panel: Residual TS map of the $\mathrm{1^{o}\times1^{o}}$ region
centered at R.A. $=$ 237\fdg1583, Decl. $=$ $-$53\fdg7867 (equinox J2000.0) 
in the 3--300 GeV range, with the catalog source 2FGL J1551.3$-$5333c also
removed. The image scale of the map 
is 0\fdg02 pixel$^{-1}$. The black contours are the MOST 843 MHz radio 
contours (at square-root scale levels 0.01, 0.02, 0.05, 0.10, 0.17, 0.25, 
and 0.36 Jy beam$^{-1}$; \citealt{wg96}). The dark and green dashed circles 
mark the 2$\sigma$ error circles of the best-fit 
positions of the Kes 27 $\gamma$-ray emission and source $B$ (see the text), 
respectively.  The latter is possibly associated with 
a background candidate AGN  (marked by the green cross; \citealt{css+2008}).}
\label{fig:hmap}
\end{figure}

\begin{figure}
\epsscale{1}
\plotone{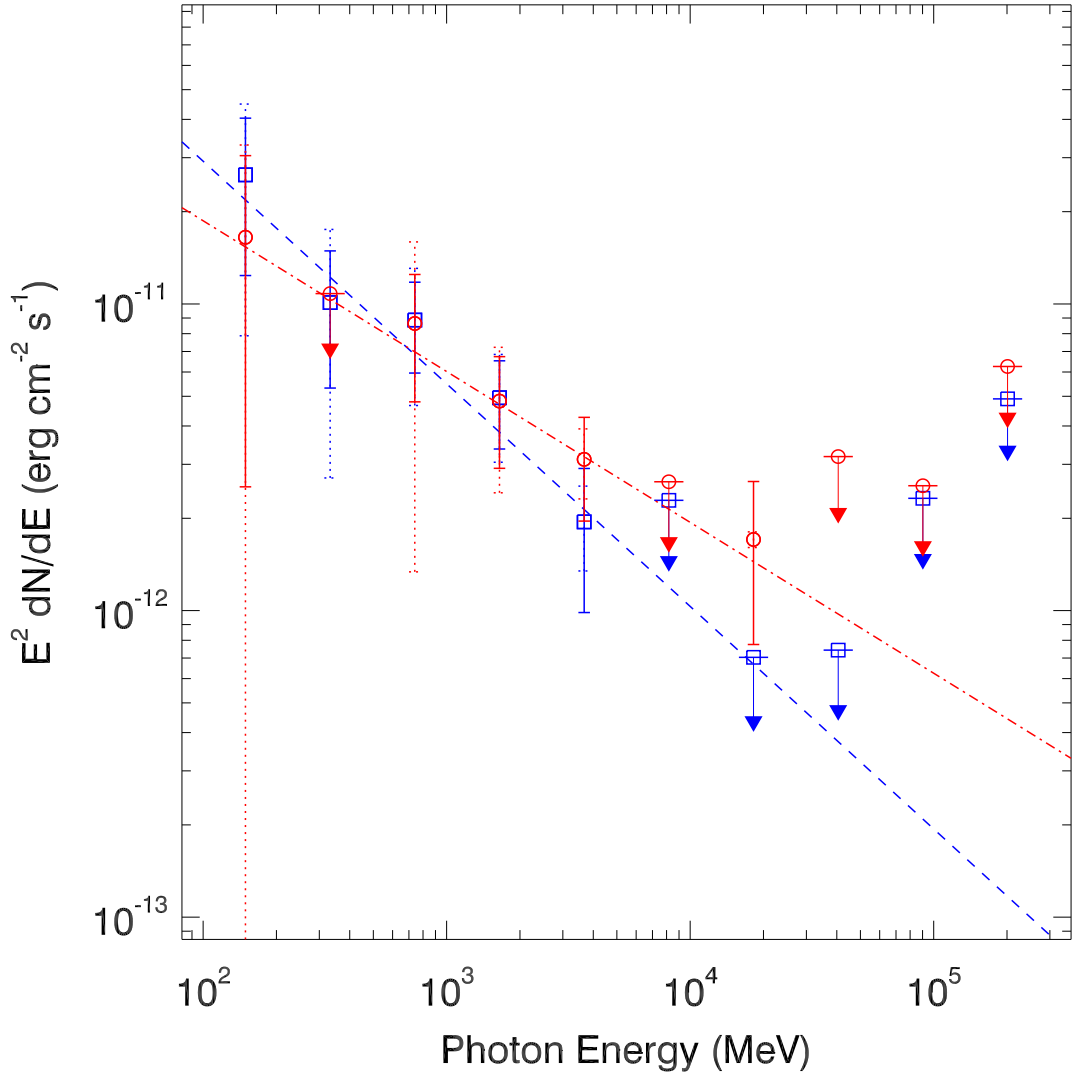}
\caption{\fermi\ \gr\ spectra and power-law fits of
Kes 27 (red circles and dash dotted line) and source $B$ 
(blue squares and dashed line), obtained from maximum likelihood 
analysis. The statistical and systematic uncertainties are shown as solid and dotted
bars, respectively.}
\label{fig:spectra}
\end{figure}

However, a high proton index can be naturally expected if the accelerated 
protons experience a diffusion process before they bombard the surrounding 
dense gas. The centroid of the $\gamma$-ray TS map is essentially located 
outside the southeastern boundary (see Figure~\ref{fig:hmap}) and appears
to be coincident with the HI cloud on the southeast 
\citep{mgd+2001}. Therefore, the adjacent cloud may be `illuminated' by 
the protons escaping from the SNR.
A convenient algorithm has been established for such a bombardment 
for the illumination of an adjacent cloud by the diffusive energetic 
protons escaping from an expanding SNR shock front \citep{lc2010}. 
In this scenario, at a given position for the dense cloud outside the SNR, 
the proton spectrum is obtained by accumulating all the contribution of 
the escaping protons throughout the entire history of the SNR expansion. 
When this accumulative collection of diffusive protons collides with 
the nearby cloud, the $\pi^{0}$-decay $\gamma$-rays emanate. 
The power-law index of the escaping protons will be higher than 
$\alpha_p$ and approaches $\alpha_p+\delta$ \citep{aa96}, 
where $\delta$ is the power-law index of diffusion coefficient $D(E_p)$ 
($D(E_p)\propto E_p^{\delta}$, and $\delta$ = 0.3--0.7; 
e.g., \citealt{ber+90}). 
We hereby refer to \citet{lc2010} and the references therein for details 
of the model.

\begin{figure}
\epsscale{0.8}
\plotone{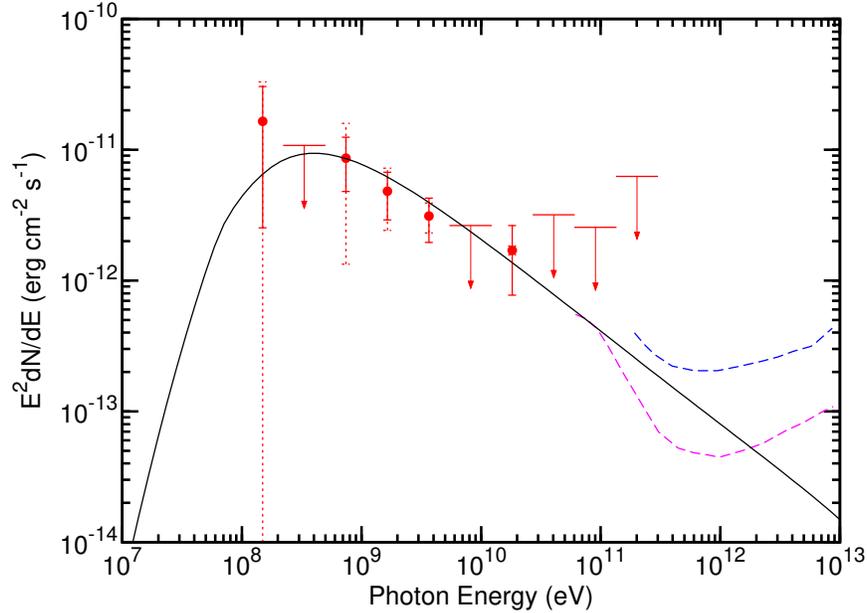}
\caption{\fermi\ \gr\ spectrum of Kes 27. The statistical and systematic uncertainties are shown as solid and dotted
bars, respectively.
The hadronic model (solid curve; \citealt{lc2010}) can reproduce 
the observed emission. The blue and pink dashed curves indicate 
the sensitivity limits of HESS and CTA, respectively.}
\label{fig:sed}
\end{figure}

In our calculation using the latter model, an age of $\sim 8000$~yr is
adopted for SNR Kes~27. At a distance of 4.3 kpc, the SNR is $\sim13$~pc 
in radius and the HI cloud is thus assumed to be 
at $R_{\rm cl}\approx13$~pc away from the center of the SNR. Other parameters 
used in the calculation were the fraction of the explosion energy converted 
into the accelerated protons $\eta = 0.1$ \citep{be1987}, 
the spectral index for the energy distribution of the protons 
$\alpha_p = 2.2$ (e.g., \citealt{giu+2010}), and the correction factor of 
slow diffusion around the SNR $\chi = 0.1$ \citep{fuj+09}. The 
parameters $\chi$ and $\delta$ determine 
the diffusion coefficient and thus the diffusion radius in the model. 
In the calculation to fit the observed $\gamma$-ray spectrum, 
$\delta = 0.4$ and a total cloud mass of 
$M_{\rm cl}\sim 200 E_{51}^{-1} M_{\odot}$ are required. The model spectrum 
is shown in Figure~\ref{fig:sed}. 
The model cloud mass is consistent with the observation. Actually, 
according to \citet{mgd+2001}, the bulk (including the core) of the HI 
cloud on the southeast appears to be outside the SNR’s edge. The hot gas 
density is $\ga 2$\,cm$^{-3}$ in region E1 \citep{css+2008}, and may be no 
higher than the average gas density of the cloud, since this region is likely 
in the outer part of the cloud. Therefore, the cloud, $\sim0.1$ degree in 
angular radius, may thus have a mass $\ga 120 M_{\odot}$.
We note that our model spectrum is slightly lower than 
the sensitivity limit of HESS (\citealt{hess}; see Figure 4), which may 
explain the non-detection of Kes 27 at TeV energies in the HESS Galactic 
plane survey \citep{car+13}. Considering the much improved sensitivity of 
the Cherenkov Telescope Array (CTA), the Fermi detection of Kes 27 can be 
confirmed by CTA observations, and the model used here may also be tested 
or constrained.

\acknowledgments

We gratefully thank anonymous referee for very constructive suggestions.
This research was supported by Shanghai Natural Science Foundation for
Youth (13ZR1464400), 
National Natural Science Foundation of China (11373055),
and the Strategic Priority Research Program ``The Emergence of 
Cosmological Structures" of the Chinese Academy of Sciences 
(Grant No. XDB09000000). Y.C. acknowledges support from 
National Natural Science Foundation of China (11233001), 
the 973 Program (Grant 2015CB857100), 
and the Ph.D. Programs Foundation of the Educational
Ministry of China (Grant 20120091110048). 
Z.W. is a Research Fellow of the One-Hundred-Talents project of 
Chinese Academy of Sciences.

\clearpage

\clearpage

\clearpage

\clearpage

\begin{deluxetable}{lcc}
\tablecaption{\fermi/LAT flux measurements of source A and B in the Kes 27 region}
\tablewidth{0pt}
\startdata
\hline
\hline
$E$ & $E^2dN(E)/dE$ (A) & $E^2dN(E)/dE$ (B) \\
(GeV) & (10$^{-12}$ erg cm$^{-2}$ s$^{-1}$) & (10$^{-12}$ erg cm$^{-2}$ s$^{-1}$) \\
\hline
0.15 & 16.5$\pm$14.0$\pm$16.5 & 26.4$\pm$14.0$\pm$18.5 \\
0.33 & 10.8 & 10.1$\pm$4.8$\pm$7.4 \\
0.74 & 8.6$\pm$3.8$\pm$7.3 & 8.9$\pm$2.9$\pm$4.2 \\
1.65 & 4.8$\pm$1.9$\pm$2.4  & 4.9$\pm$1.6$\pm$1.9 \\
3.67 & 3.1$\pm$1.2$\pm$0.8 & 1.9$\pm$1.0$\pm$0.6 \\
8.17 & 2.6 & 2.3 \\
18.20 & 1.7$\pm$0.9$\pm$0.1 & 0.7 \\
40.54 & 3.2 & 0.8 \\
90.27 & 2.6 & 2.3 \\
201.03 & 6.3 & 4.9 \\
\enddata
\tablecomments{The first uncertainties are statistical uncertainties. The second uncertainties are systematic uncertainties introduced by the Galactic diffuse emission. The 5--10\% systematic uncertainties introduced by the LAT effective area are not listed here. Fluxes without uncertainties are the 95$\%$ upper limits.}
\label{tab:tab1}
\end{deluxetable}


\end{document}